# Performance Analysis of Boron Nitride Embedded Armchair Graphene Nanoribbon MOSFET with Stone Wales Defects

Anuja Chanana, Amretashis Sengupta and Santanu Mahapatra

*Nano Scale Device Research Laboratory Department of Electronic Systems Engineering, Indian Institute of Science, Bangalore 560 012, India*

We study the performance of a hybrid Graphene-Boron Nitride (GNR-BN) armchair nanoribbon (a-GNR-BN) MOSFET at its ballistic transport limit. We consider three geometric configurations 3p, 3p+1 and 3p+2 of a-GNR-BN with BN atoms embedded on both sides (2, 4 and 6 BN on each side) on the GNR. The material properties like band gap, effective mass and density of states of these H-passivated structures have been evaluated using the Density Functional Theory (DFT). Using these material parameters, self-consistent Poisson-Schrodinger simulations are carried out under the Non Equilibrium Green's Function (NEGF) formalism to calculate the ballistic MOSFET device characteristics. For a hybrid nanoribbon of width ~5 nm, the simulated ON current is found to be in the range 276 µA - 291 µA with an ON/OFF ratio 7.1 x $10^6$ - 7.4 x $10^6$ for a $V_{DD}$=0.68 V corresponds to 10 nm technology node. We further study the impact of randomly distributed Stone Wales (SW) defects in these hybrid structures and only 2.52% degradation of ON current is observed for SW defect density of 6.35%.

## I. INTRODUCTION

After the successful isolation of graphene[1] from the bulk graphite, it has become a major material for electronic applications owing to its planar structure and novel properties [2-3] like high electron mobility, high thermal conductivity, flexibility and optical transparency. However the zero band gap of graphene [4-5] makes the graphene FET unsuitable for logic applications. In this regard a possible solution is the lateral confinement of carriers in a graphene nanoribbon (GNR) [6] to open a band gap. This band gap depends on the width and chirality of the GNR[7] which makes it a possible choice as a channel material for MOSFET [8]. However, the band gap of GNR becomes quite small of the order of few meV with an increase in the width of the nanoribbon [4-5] beyond 4 nm. Moreover, their fabrication with considerable accuracy is itself a significant challenge [9-12]. Due to the structural similarity between hexagonal monolayer boron nitride (BN) and graphene, BN attracts a great interest as a suitable dopant/ embedding material [13-22] for graphene. BN nanoribbons (BNNR) [23-25] demonstrate much higher band gap than that of GNR. Boron Nitride [26-28] embedded GNR leads to an enhancement of a band gap [29-31] of pure GNR and yet preserving the low values electron effective mass in GNR to some extent. Such hybrid structures of a-GNR-BN have been successfully fabricated [32-33] and thus appear to be potential channel material for future nanoscale MOSFET. Though several studies have been made on the material properties of hybrid a-GNR-BN, to our best knowledge, there is no report on the performance analysis of a MOSFET using them as channel material.

Here we report the performance limit of a hybrid Graphene-Boron Nitride [34-36] armchair nanoribbon (a-GNR-BN) MOSFET in the context of 10 nm technology node [37]. We consider three geometric configurations 3p, 3p+1 and 3p+2 of a-GNR-BN with BN atoms embedded on both sides (2, 4 and 6 BN on each side) of the GNR. The widths made of total 42, 43 and 44 atoms are considered for the present study. Three substructures are realized for a particular width of hybrid-a-GNR-BN, such as 38GNR_4BN, 34GNR_8BN and 30GNR_12BN for 42 a-GNR-BN; 39GNR_4BN, and 35GNR_8BN, 31GNR_12BN for 43 a-GNR-BN and 40GNR_4BN, 36GNR_8BN, 32GNR_12BN for 44 a-GNR-BN. H-passivation is considered to reduce contribution from edge states. The material properties like band gap and effective mass has been evaluated using the Density Functional Theory (DFT). Using these material parameters, self-consistent solution of Poisson-Schrodinger equation were carried out under the Non Equilibrium Green's Function (NEGF) formalism to calculate the ballistic MOS device characteristics. We study the various output characteristics of the hybrid a-GNR-BN MOSFET like $I_D$-$V_D$, $I_D$-$V_G$, gm-$V_G$, $V_G$-cutoff frequency, $I_{ON}/I_{OFF}$, Drain induced Barrier Lowering (DIBL) and Subthreshold Slope(SS). Since the channel length is 10 nm, the transport is assumed to be purely ballistic in the devices. A common defect observed in the GNR [38-40] structure is the Stone-Wales (SW) defect. The effects of the SW has been studied extensively both theoretically [41-44] and experimentally [45-46]. Since hexagonal BN is also $sp^2$ bonded material, SW is also observed in it [47-49]. In our work we also study the impact of SW defect of the ballistic device performance of hybrid a-GNR-BN MOSFET.

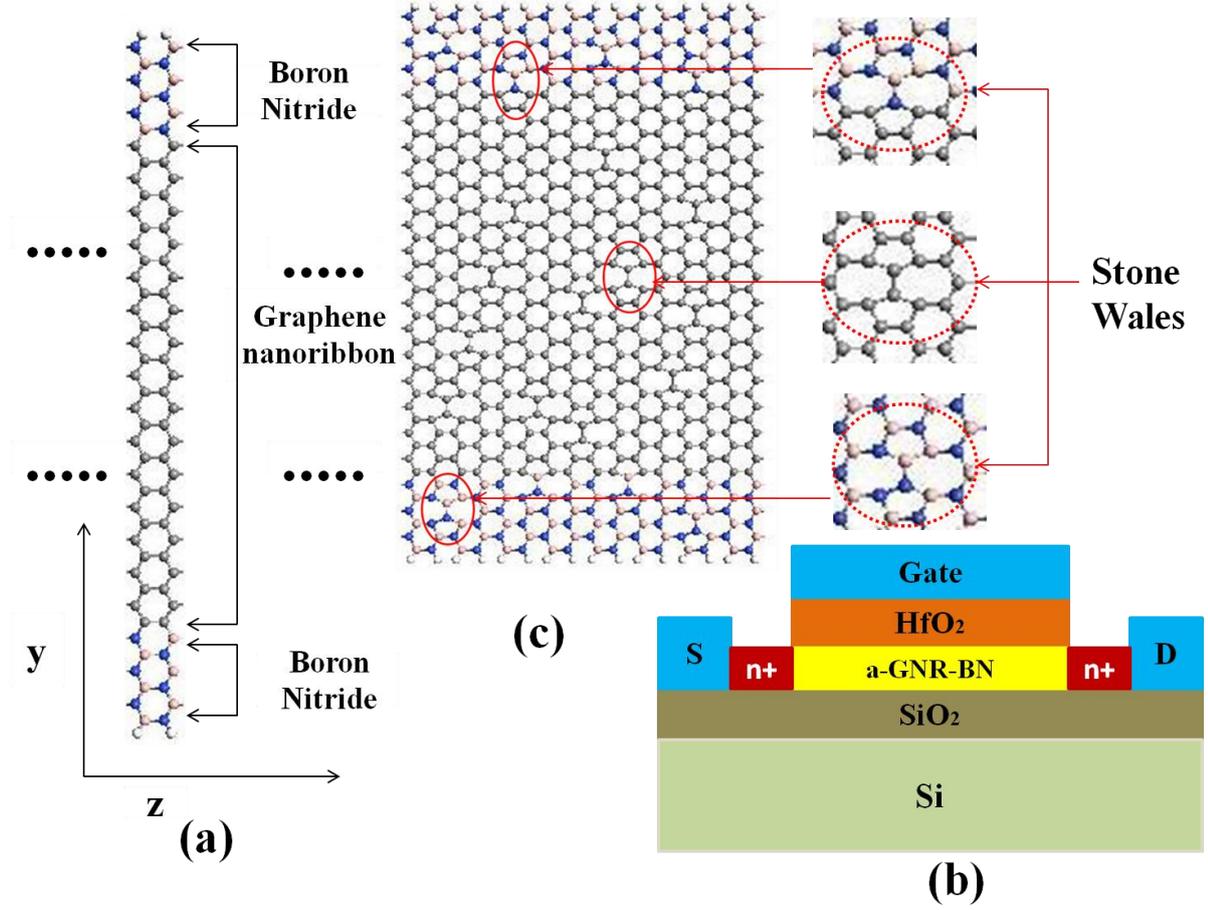

FIG. 1.(a) Structure of hybrid-a-30GNR12BN (z-axis is the transport direction) We can see here the C atoms which are perturbed by adjacent B and N atoms are anti symmetric on the opposite side of the nanoribbon, (b) Device schematic (not to scale) of the hybrid-a-GNR-BN considered in our studies. (c) Structure of defected supercell consisting of 20 Stone Wales (SW) MOSFET. Zoomed view of each SW defects in pure graphene, BNNR and at the interface.

## II. METHODOLOGY

Fig. 1(a) shows the structure of the hybrid a-30GNR-12BN. The transport direction is z-direction. The schematic cross sectional view of the MOSFET is shown in Fig.1 (b). The hybrid a-GNR-BN of length 10 nm is used as the 2D channel material; the channel width varies depending upon hybrid a-GNR-BN configuration, which is 5.05, 5.17 and 5.29 nm for 42, 43 and 44 atoms. This 2-D channel is placed over a $SiO_2$/Si substrate. $HfO_2$ is taken as the gate dielectric having a thickness of 2.5 nm. Highly doped ($10^{20}$ /cm$^3$) n$^{++}$ regions serve as the source and drain contacts for the NMOSFET.

Density Functional Theory (DFT) calculations are performed to evaluate material properties of a-GNR-BN using QuantumWise ATK [50]. The Localized Density Approximation (LDA) exchange correlation with a Double Zeta Polarized (DZP) basis is used with mesh cut-off energy of 75 Ha [51]. We use Troullier-Martins type norm-conserving pseudopotential sets in ATK (NC-FHI [z=1] DZP for Hydrogen, NC-FHI [z=4] DZP for Carbon, NC-FHI [z=3] DZP for Boron and NC-FHI [z=5] DZP for Nitrogen). The Pulay-mixer algorithm is employed as iteration control parameter with tolerance value of $10^{-5}$. The maximum number of iteration step is set to 100. We use a 1x1x16 Monkhorst-Pack k-grid mesh for our simulations [52]. The material properties show a negligible change when the grid points in the transport z-direction are increased. For the DFT calculations if we use Gradual Gradient Approximation (GGA) as the exchange correlation, the band gap and the effective mass showed a minimal change.

Stone-Wales (SW) defect is a kind of point defect defined by the reconstruction of graphene lattice by formation of non-hexagonal rings, and is comprised of two pairs of five-membered and seven-membered rings (5-7-7-5) [38-40]. The structure chosen for SW defect analysis is hybrid a-42GNR-BN of 3p configuration, which consists of 30 GNR atoms and 12 BNNR atoms. A supercell of length 3.23 nm is realized and henceforth SW defects are introduced in this structure. Fig. 1(c) shows the supercell having 20 SW defects. One can observe from Fig. 1(c) that the SW defects are distributed randomly in the GNR part, BNNR part and at their interface. As the supercell contains large number of atoms, Slater-Koster method [53] instead of DFT is used to evaluate the material properties of the supercell for better convergence. For this we use the DFTB (CP2K) non self consistent basis set with a mesh cut off energy of 10 Ha [51]. The Possion solver is FFT and the maximum range of interaction is 10 Angstroms. A 1x1x16 Monkhorst-Pack k-grid mesh is used for the simulations [52]. We later show that the bandstructure of 42 hybrid a-GNR-BN calculated using both DFT and Slater Koster shows almost same nature of band structure especially at conduction band minima and valence band maxima. Therefore it is expected that supercell band structure simulation using Slater Koster method will yield the value band-gap and effective masses comparable to the DFT technique.

We obtain the bandgap and effective mass of different nanoribbons using the above mentioned methodology, which are then used in NEGF simulator [54-55] to calculate the fully ballistic transistor performance analysis. In NEGF formalism, self-energy matrices for the source and drain contacts ($\Sigma_S$ and $\Sigma_D$) are used to construct the Green's function $G$ as

$$G(E) = [EI - H - \Sigma_S - \Sigma_D]^{-1} \qquad (1)$$

where $I$ is the identity matrix. Since the transport assumed is purely ballistic, so no scattering matrix has been included in the Green's function [56]. Eq. (1) can be used to evaluate parameters like the broadening matrices $\Gamma_S$ and $\Gamma_D$ and the spectral densities $A_S$ and $A_D$ defined by the following relations:

$$\Gamma_{S,D} = i[\Sigma_{S,D} - \Sigma^{\dagger}_{D,S}] \qquad (2)$$

$$A_{S,D} = G(E)\Gamma_{S,D}G^{\dagger}(E) \qquad (3)$$

The density matrix [$\mathbb{R}$] used to solve the Poisson equation is given by

$$[\mathbb{R}] = \int_{-\infty}^{\infty} \frac{dE}{2\pi}[A(E_{k,x})]f_0(E_{k,x} - \eta) \qquad (4)$$

where A($E_{k,x}$) is the spectral density matrix, $E_{k,x}$ the energy of the conducting level, η is the chemical potential of the contacts and $f_0$ is the Fermi function.

For the Poisson solver, we use finite difference methods as similar to Guo *et al.*[57] and Ren [58]. The transmission matrix *T (E)* is calculated as

$$T(E) = Trace[A_S\Gamma_D] = Trace[A_D\Gamma_S] \qquad (5)$$

And thus the ballistic drain current is calculated as

$$I_D = \left(\frac{4e}{h}\right)\int_{-\infty}^{+\infty} T(E)\left[f_S\left(E_{k,x} - \eta_S\right) - f_D\left(E_{k,x} - \eta_D\right)\right]dE \qquad (6)$$

In Eqn. (6), *e* is the electronic charge, *h* is the Planck's constant, $f_S$ and $f_D$ are the Fermi functions in the source and drain contacts. $\eta_S$ and $\eta_D$ are the source and drain chemical potentials respectively. The spin degeneracy and valley degeneracy in nanoribbon accounts for a factor of 4 in the above equation. A complete ballistic transport is depicted by Eqn. (6) which can be used for ultra short channel lengths of 10 nm.

**III RESULTS & DISCUSSIONS:**

From the ab-initio calculations we find that all the nanoribbons show a direct band gap at Γ point and Z point of the Brillouin zone. As shown in Fig. 2(a) the highest band gap 0.369 eV is obtained for the hybrid a-42GNR-BN, a 3p configuration made of a-12BNNR and a-30GNR. Fig. 2(b)-(d) also depict bandstructure of 42GNR, 30 GNR, 12BNNR separately. As it can be seen from the Fig.2 (a) and Fig.2 (b), the band gap of BN doped structure is higher as compared to the pure GNR. In Fig 2 (e), bandstructure of hybrid a-42GNR-BN is obtained using DFT and Slater Koster and it is observed that band gap is same for both the methods.

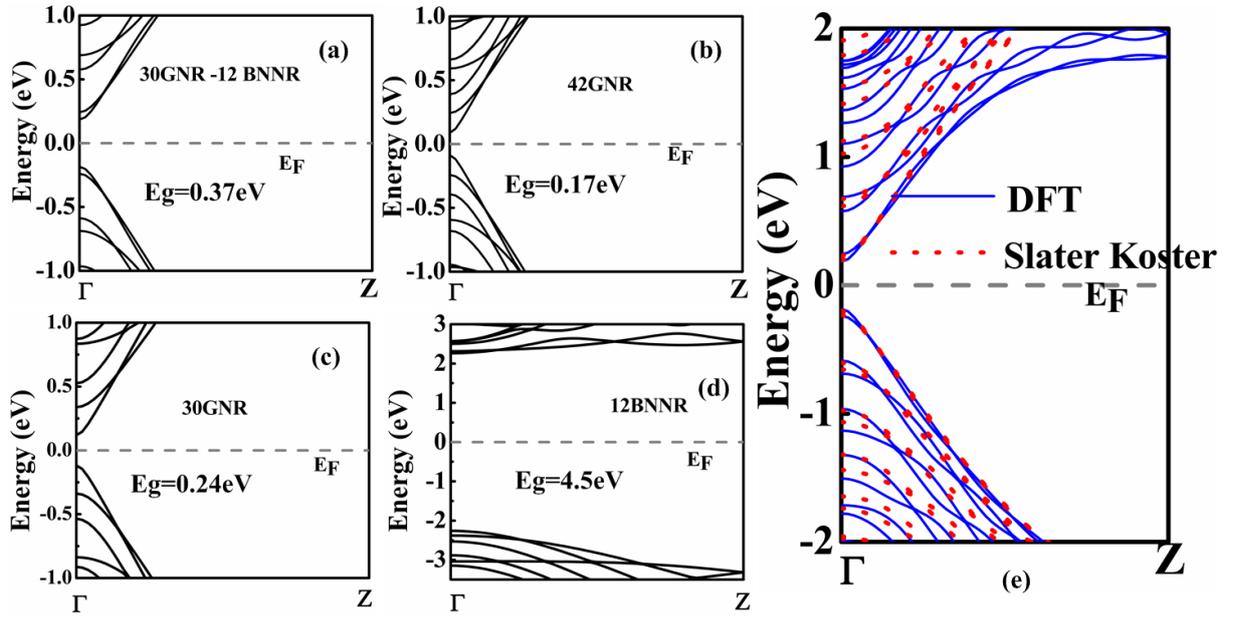

FIG.2. DFT calculated band structures of the hybrid-a-30GNR-12BN (a), 42GNR (b), 30GNR(c) and 12BNNR (d). (e) Comparison of band structures evaluated for hybrid a-30GNR-12BN using DFT and Slater Koster method.

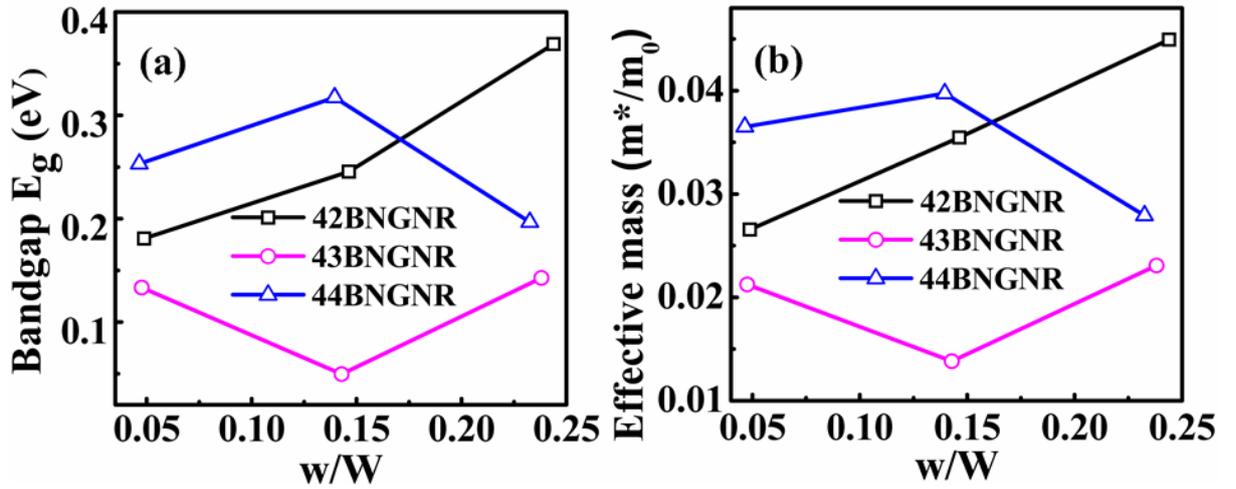

FIG.3. Variation of Band gap (a) and Effective mass (b) for the widths 42, 43, 44 w.r.t ratio of BN width (w) to the whole nanoribbon width(W).

Fig. 3(a) and (b) show the nature of band gap and effective mass for the hybrid armchair nanoribbon made of 42, 43 and 44 GNR and BNNR, each of width 5.05 nm, 5.17 nm and 5.23 nm respectively, with increasing BN concentration. For each width, there are substructures that have 3p, 3p+1 and 3p+2 configurations of each a-GNR and a-BNNR. For the width 42 hybrid-a GNR-BN (5.05 nm) there are 38GNR(3p+2)_4BNNR(3p+1), 34GNR(3p+1)_8BNNR(3p+2) and 30GNR(3p)_12BNNR(3p) substructures. It can be observed from the Fig.3 (a) and Fig. 3(b) that the 3p configuration of width 42 hybrid-a-GNR-BN which contains maximum BN concentration (30GNR_12BNNR) has the highest bad gap 0.369 and highest effective mass 0.045 $m_0$.

We also find that for this configuration (42 hybrid a-GNR-BN) when the BN doping increases in the multiples of 2 on both sides, the band gap and effective mass increase linearly with the increasing BN concentration. For 3p+1 configuration (43hybrid a GNR-BN) it first decreases and then increases and for 3p+2 configurations (44hybrid a-GNR-BN) the nature is vice versa. For an even count of hybrid a-GNR-BN (total number of atoms in the hybrid nanoribbon i.e., 42 and 44), the hierarchy of the band gap is $E_{3p+2}<E_{3p+1}<E_{3p}$ which is very well in agreement with the previous results [29, 30]. For an odd count of hybrid a-GNR-BN (total number of atoms in the hybrid nanoribbon i.e., 43) the hierarchy observed is $E_{3p+1}<E_{3p+2}<E_{3p}$ which resembles to the hierarchy of pure GNR.

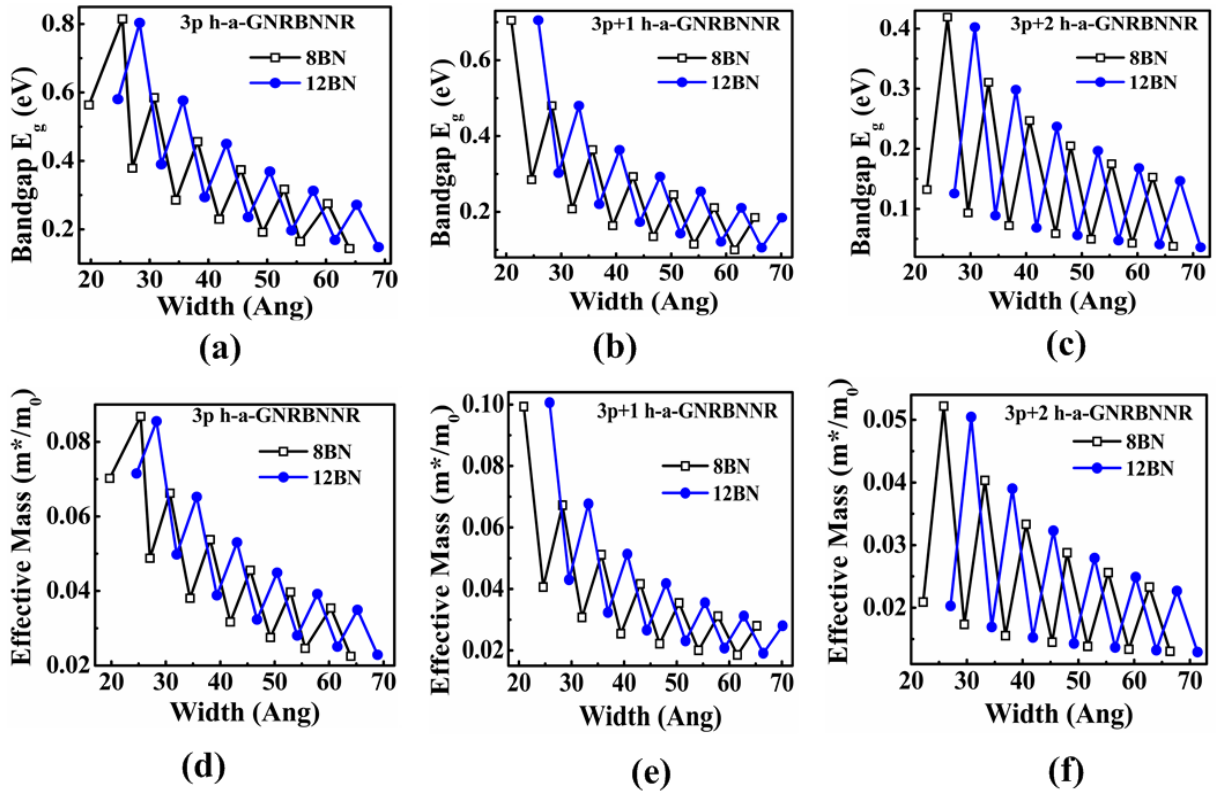

FIG.4. Band Gap vs. width of hybrid-a GNR-BN for 8BN and 12 BN doping respectively of (a) 3p, (b), 3p+1 and (c) 3p+2 configuration of graphene. Effective mass vs. width of hybrid-a GNR-BN for 8BN and 12 BN doping respectively of (a) 3p, (b), 3p+1 and (c) 3p+2 configuration of graphene.

To confirm this nature we carry some simulations for structures, which are made by increasing GNR atoms for each configuration (3p, 3p+1, 3p+2) and BN doping of 2, 4 and 6 atoms on each side. While making these structures the count of the total atoms in the nanoribbon made of both graphene and BN goes odd and even alternatively. The band gap and effective mass obtained for a doping of 8BN and 12BN in the hybrid nanoribbon for all the 3 configurations are shown in Fig. 4. As we can see the band gap and effective mass decrease showing a decaying zigzag nature as the width of GNR increases and keeping the BN width same on both sides. The nanoribbon containing an odd number of atoms has a low band gap and effective mass as compared to ones which have an even number of atoms.

The partial density of states (PDOS) of pure and hybrid nanoribbon is also shown in Fig. 5(a) and 5(b). The DOS for s, p, and d orbital have been shown separately. One can see that the contribution of the p-orbital is much more in PDOS as compared to the s and d orbital.

The PDOS for each atom i.e., carbon, boron and nitrogen in the hybrid nanoribbon is evaluated separately and it is observed that the p orbital of carbon atom accounts for the maximum PDOS as shown in Fig.5. (b).

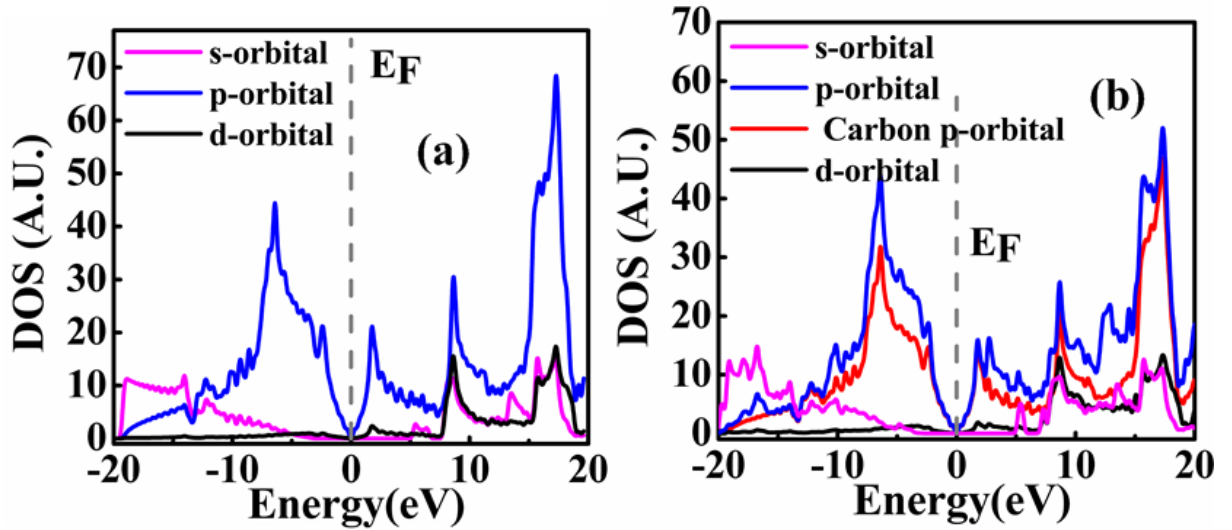

FIG.5. (a) PDOS of pure 42GNR, (b) PDOS of hybrid a-30GNR12BN.

Using the calculated material properties, we solve the Poisson-Schrödinger equation of our system self-consistently under the NEGF formalism [56, 59] as discussed in section II. The simulated output characteristics of the hybrid armchair nanoribbon (42 hybrid a-GNBNR) based MOSFET is shown in Fig.6 (a)-(d). From the results in Fig. 6(a), we observe the value of ON current 290.80 µA, 283.10 µA and 275.7 µA at Vg=0.68 V for the substructures hybrid-a-38GNR_4BN, hybrid a-34GNR_8BN, hybrid a-30GNR_12BN respectively. As we can see the substructure with the lowest effective mass has the highest value of ON current. For the above mentioned substructures, the ON/OFF ratio is calculated as $7.12 \times 10^6$, $7.25 \times 10^6$ and $7.38 \times 10^6$. The DIBL varies in the range 11.20-11.90 mV/V which is quite less and can favor the use of hybrid nanoribbon structures as MOSFET channel materials. The SS varies from 62.38 – 62.129 mV/decade. From Fig. 4(c) and Fig. (d), the peak transconductance has been found as 525.02 µS, 524.45 µS and 523.95 µS at Vg=0.42 and the maximum values of cut off frequency ($f_T = g_m/2C_G$) obtained is 3.6 THz, 3.6 THz and 3.4 THz. These values also correspond to the substructures as cited before. The gate capacitance for the MOSFET (Fig. 1(c)) is calculated as $4427 \times 10^{-9}$ pF/m.

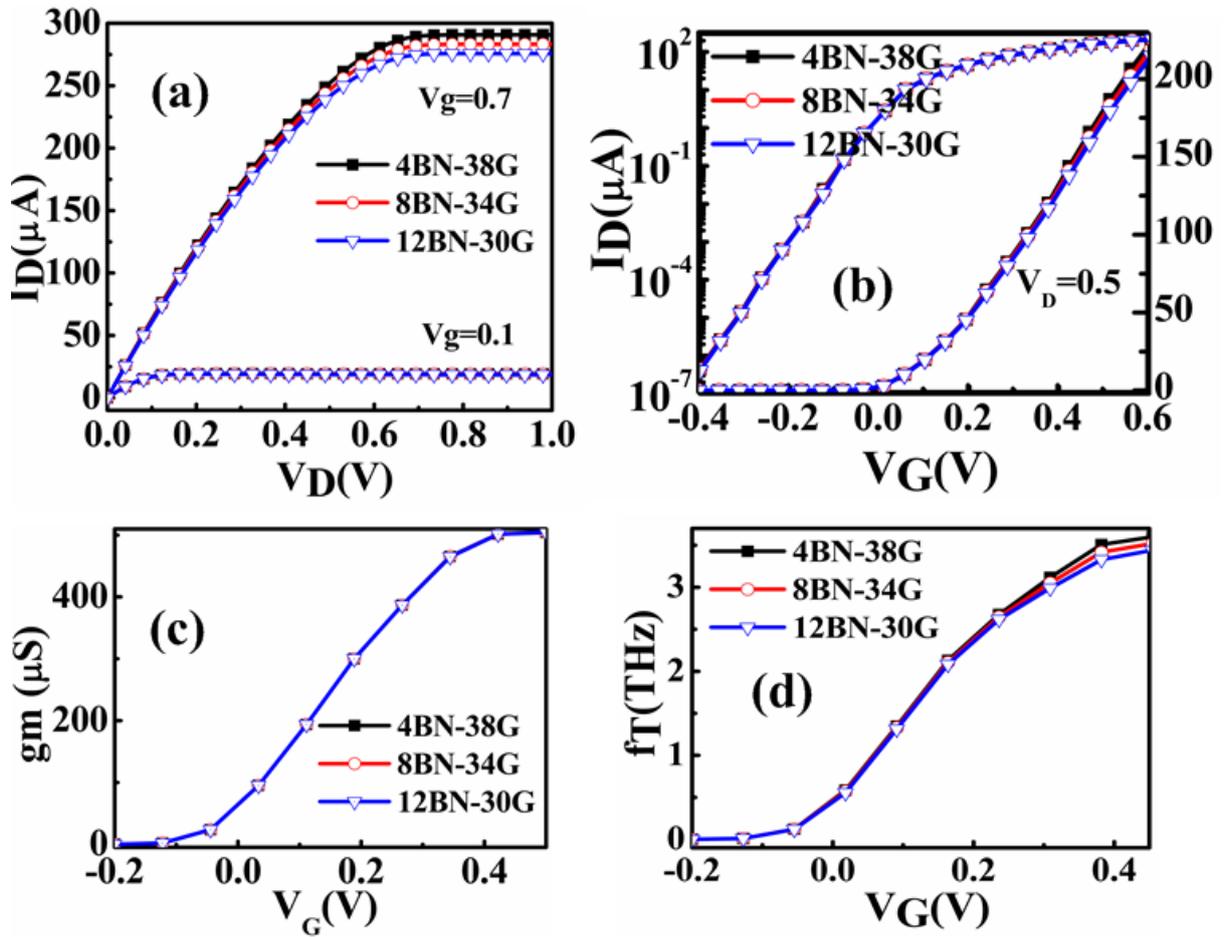

FIG.6. Simulated device characteristics of hybrid-a-30GNR12BN (a) $I_D$-$V_D$ evaluated at $V_G$=0.1 and $V_G$=0.7. (b) $I_D$-$V_G$ calculated at $V_D$=0.5V. Simulated $g_m$-$V_G$ (c) and cut-off frequency (d) for hybrid-a 30GNR12BN based MOSFET.

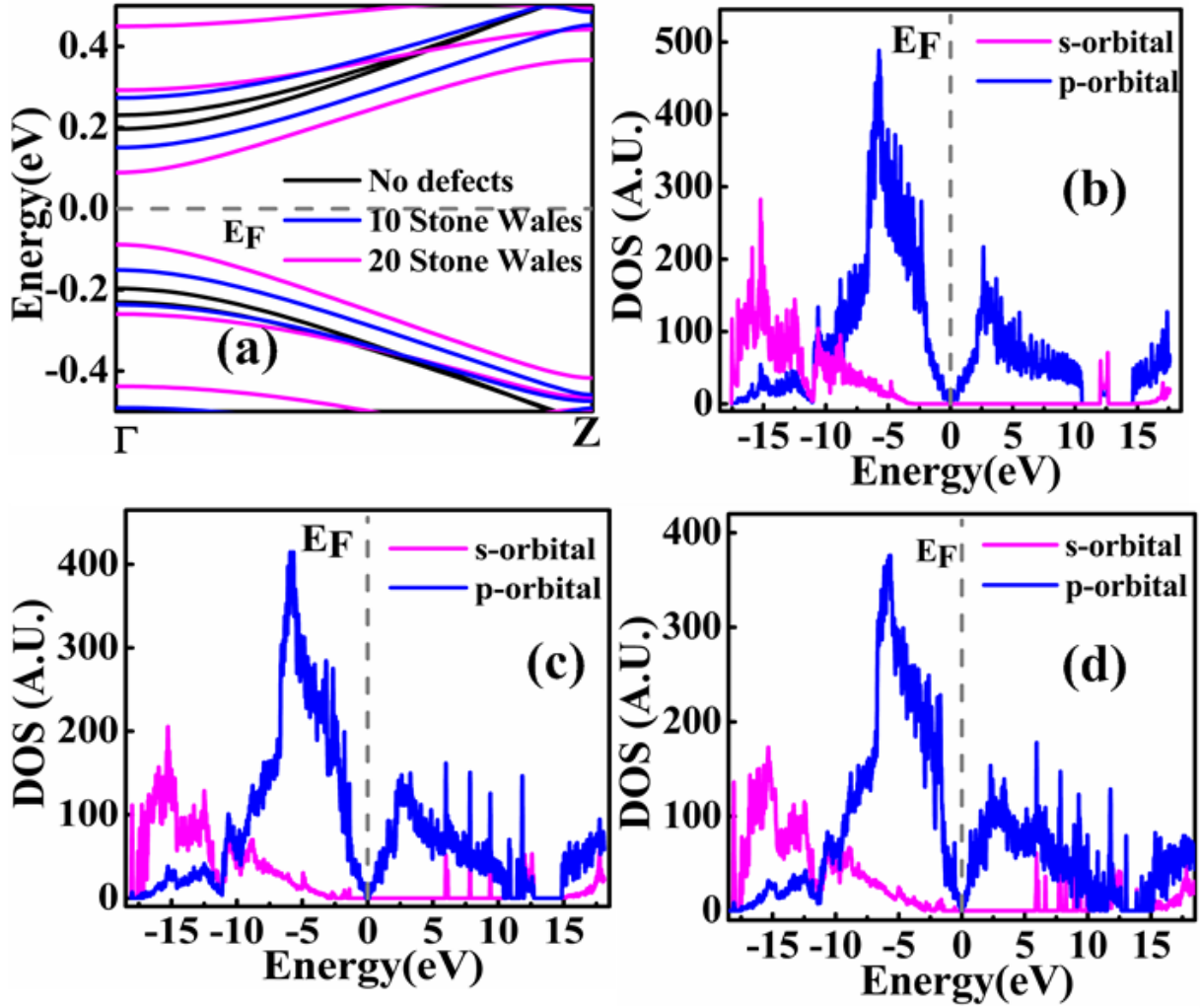

FIG.7. (a) Comparison of band structures of pure supercell with 10 SW and 20 SW affected supercell (supercell is made by repeating the hybrid-a-30GNR-12BN structure). PDOS of pure supercell (b) 10 SW (c) and 20 SW (d) affected supercell

Fig. 7(a) compares the bandstructure of supercell of hybrid-a 42GNR-BN (Fig. 1(b)) with and without the Stone-Wales defect. A total of 3 structures were realized (a) pure (without SW) supercell (b) with 10 SW defects (c) with 20 SW defects (Fig.1. (b)). The projected PDOS for the pure and defected super cell with 10 and 20 SW has been shown in the Fig. 7 (b)-(d). It is observed that the PDOS is decreasing when the number of defects increases. It is the least when the SW defect is 20. We can also see that the contribution of p orbital is larger in comparison to the s orbital. Sharp peaks can be observed in the energy range 5 eV – 12.5 eV by the p-orbital and for the s-orbital the DOS is smoothened in the energy range -20 eV – - 10 eV in the defected structure as compared to pure.

Table 1 shows the band gap, effective mass, $I_{ON}$ for pure and defected structure. Here we can observe that the when the number of defects are increasing, the band gap and effective mass decreases as a result of which the ON current increases. There is also a reduction in the intrinsic delay time. As we can observe from the Table 1, by introducing 10 SW defects, the band gap and effective mass decrease by ~ 23.3% and ~7%. On the other hand the decrease observed is ~58% and ~38% when the number of defects is increased to 20 SW. The change in $I_{ON}$ and intrinsic delay time is maximum to ~0.5% and ~3% for 10SW and 20SW respectively. Here we conclude that the 6.35% of defect density (20SW among 315 honeycombs in supercell) leads to significant modification in band gap and effective mass of the supercell but a negligible change is observed in the $I_{ON}$ and delay time for the device.

## Table I

| Structure | Band gap (eV) | Effective Mass($m^*/m_0$) | $I_{ON}$ (µA) | Tau (ps) |
|---|---|---|---|---|
| Pure supercell | 0.39321 | 0.0482577 | 97.31 | 0.0318 |
| Defected Supercell (10SW) | 0.30162 | 0.0448896 | 97.82 | 0.0317 |
| Defected Supercell (20SW) | 0.17706 | 0.0298179 | 100.27 | 0.031 |

**IV. CONCLUSION:**

In the paper, we study the performance of hybrid armchair graphene BN nanoribbons as a channel material in the 10 nm technology node for MOSFET. The material properties are calculated using DFT for unit cell of hybrid nanoribbon and Slater Koster for the supercell of hybrid nanoribbon. The device characteristics are evaluated using the self consistent Poisson-Schrödinger solutions performed under the NEGF formalism. The hybrid nanoribbon shows a higher band gap and effective mass gap as compared to the pure nanoribbon. The ballistic device characteristics such as ON current, ON/OFF ratio, transconductance, DIBL etc. depict a good performance which makes the hybrid nanoribbon as a potential candidate for 10 nm technology node. Among all the configurations of hybrid nanoribbon, the 3p configurations showed the maximum band gap and best performance in terms of MOSFET characteristics. The effects of Stone-Wales (SW) defects were further studied and it is observed that with the increase in number of SW the band gap and effective mass decreases and the ON current increases.

.